\documentclass[a4paper,twocolumn,fleqn]{article} 
\usepackage{cite}             
\usepackage{amsmath}
\usepackage{textcomp}
\usepackage{amssymb}

\usepackage{caption}
\usepackage{upgreek}
\usepackage{xcolor}
\usepackage{graphicx}

\usepackage{pfr_itc23}        

\begin{document}

\title{Coulomb collisional effects on high energy particles
in the presence of driftwave turbulence}

\author{B.~Huang, Y.~Nishimura, and C.~Z.~Cheng}

\affiliation{
Institute of Space, Astrophysical, and Plasma Sciences \\
National Cheng Kung University, Tainan 70101, Taiwan}

\date{(Received 21 November 2013)}

\email{nishimura@pssc.ncku.edu.tw}

\begin{abstract}
High energy particles' behavior including fusion born alpha particles in an ITER like tokamak
in the presence of background driftwave turbulence
is investigated by an orbit following calculation.
The background turbulence is given by the toroidal driftwave eigenmode
combined with a random number generator.
The transport level is reduced as the particle energy increase;
the widths of the guiding center islands produced by the passing particles are
inverse proportional to the square root of parallel velocities. On the other hand,
the trapped particles are sensitive to $E \times B$ drift at the banana tips
whose radial displacement is larger for lower energy particles.
Coulomb collisional effects are incorporated
which modifies the transport process of the trapped high energy particles
whose radial excursion resides in limited radial domains without collisions.
\end{abstract}

\keywords{\normalsize orbit following particle calculation, driftwave turbulence, 
guiding center model, Coulomb collisions, alpha particles, ballooning mode, tokamaks}

\AbstNum{P1-19}
\CorrespondingAuthor{Yasutaro~NISHIMURA}
\CorrespondingAffiliation{National Cheng Kung University}
\PostalAddress{1 University Rd., Tainan 70101, Taiwan}
\Tel{+886.5-275-7575}
\Fax{+886.5-275-7575}
\Topic{1}

\maketitle

\begin{normalsize}

\newpage
\section{Introduction}
For a self-ignited tokamak plasma, confinement of alpha particles is a crucial issue.
The alpha particles must effectively heat the bulk plasma
without a spontaneous loss. Alpha particles heat electrons first and
then ions through electron-ion thermal equilibriation.

Recent gyrokinetic simulation studies have pointed out that the 
microinstabilities strongly affect the high energy particles' 
transport\cite{est06}.
Another gyrokinetic study of ion temperature gradient turbulence suggested 
that the diffusion of the high energy particles is small\cite{zha08,whi89}.
Note that the ion temperature gradient turbulence is generated by selfconsistent particle-in-cell method
by the thermal ions, while the high energy particles' effect on the turbulence is absent.
In this work, we study transport of alpha particles in the presence
of driftwave turbulence.

We employ the guiding center orbit calculation with the toroidal driftwave model\cite{con79,che80,nis97}.
The model provides us with good tools to understand the basic mechanism of the alpha
particle transport. By our guiding center calculation, 
we are able to compare analytical aspects and numerical work more closely. 
Furthermore, the model is flexible so that we can apply the simulation to large tokamak size, 
for example ITER. 

Our recent work\cite{hua13} clarified the onset condition of guiding center
stochasticity for the passing particles (guiding center islands' widths are
inverse proportional to the square root of parallel velocities).
On the other hand (while the resonance
in the phase velocity is possible) the radial excursion of trapped particles
is limited in finite radial domains. 

In this work, we conduct numerical experiments on collisional effects on 
the high energy particles. The small collisional effects can free the trapped
particles from the limited radial domain. 

This paper is organized as follows.
In Sec.2, we describe the computational model and present our orbit following calculation
results in Sec.3. In Sec.4, we discuss the collisional effects.
We summarize this work in Sec.5.

\section{Guiding center equation and computational model} \label{section1}
For the guiding center equation, we employ Littlejohn's guiding center Lagrangian $L$\cite{lit83}  
in the MKS unit
\begin{equation}  \label{eq:Lagrangian-Hamiltonian}
 L=q_s {\bf A}^\star \cdot  \dot{\bf X} + \frac{m_s}{q_s} \mu \dot{\phi} 
-  q_s \Phi - \mu B - \frac{1}{2} m_s v_{\parallel}^2  .
\end{equation}
 The guiding center position is given by ${\bf X}$,
$\mu$ is  magnetic moment, $\phi$ is the gyro-phase, and $v_{\parallel} $ is the parallel velocity.
The values $m_s$ and $q_s$ are particle's mass and charge, respectively.
Here, ${\bf A}^* $ is the effective vector potential,
while ${\bf B}^* = \nabla \times {\bf A}^* = {\bf B} + \nabla \times (m_s v_\| / q_s) \hat{b}$ 
is the effective magnetic field.
The unit vector along magnetic field is given by $\hat{b}$.
Here, $\Phi$ represents the electrostatic potential, 
$B$ is the magnetic field strength. 

From Eq.(\ref{eq:Lagrangian-Hamiltonian}), Euler-Lagrange equation 
gives rise to the equation of motion along with the definition of parallel velocity, 
magnetic moment, and the gyro-phase.
The guiding center equation is given by
\begin{equation} \label{eq:guidingcnterequation}
{\dot{\bf X}} = \frac{1}{B^*_{\|}} 
\left( v_{\|} {\bf B}^* + \frac{\mu }{q_s} \hat{b} \times \nabla B - \hat{b} \times {{\bf E}^*}\right) ,
\end{equation}
and the acceleration along the magnetic field by
\begin{equation} \label{eq:dot_v_parallel}
 \dot{ v_{\parallel}}  = -\frac{q_s}{m_s} \frac{1}{B^*_\|} \left(\frac{\mu }{q_s} {\bf B}^* \cdot 
\nabla B - {\bf B}^* \cdot  {\bf E}^* \right) ,
\end{equation}
where ${\bf E}^* = - \nabla \Phi - \partial_t {\bf A}$ is the effective electric field
and ${B^*_\|} = {\bf B}^* \cdot {\hat b}$.
The dot operator is for the time derivative. Hereafter, $t$ represents time.

The ballooning type electrostatic perturbation is employed 
in the orbit-following calculation\cite{con79,nis97}.
The ballooning mode posses translational invariance\cite{zak78,azu91}:
Fourier modes with different $m$ (poloidal mode number) but the same $n$ (toroidal mode number) 
are correlated. While modes 
with different poloidal/toroidal numbers are orthogonal in a cylindrical geometry, in a toroidal geometry, 
due to the toroidal curvature $1/R = (1/R_0) (1 - \varepsilon r \cos{\theta})$ ($R$ is the major radius,
$R_0$ is $R$ at the magnetic axis, $r$ and $\theta$ are minor radius and poloidal angle respectively, 
$\varepsilon = a / R_0$ is the inverse aspect ratio), 
the amplitude of neighboring poloidal modes ``$m$'' (with the same toroidal mode number ``$n$'', 
$m/n=10/10, 11/10, 12/10 ...$ ), are correlated with each other and grow at the same rate\cite{zak78,azu91}.
We describe the turbulence potential $\Phi$ by\cite{con79,che80,nis97}
\begin{eqnarray}
\Phi \left( r, \theta , \zeta, t \right) &=& \sum_{\substack{ m, n }} 
\Phi_{n} (r) \exp\left[ - \frac{\left( r-r_{mn}\right)^2}{d^2} \right] \nonumber \\
&\times&\cos\left( m \theta - n \zeta \pm {\omega}_{\star n} t + \Delta_n \right) 
\end{eqnarray}
where $\zeta$ is the toroidal angle and $r_{mn}$ is the radial location of the mode rational surfaces.
The amplitudes of $\Phi_{n} (r)$ (at the order of $10^{-3}$ of the electron temperature),
the diamagnetic frequencies ${\omega}_{\star n}$
and the phase of the mode $\Delta_n$
are given by a random number generator within physically reasonable range.
Here, $d$ is the Gaussian localization parameter\cite{nis97}.
We parameterize $m$ and $n$ so 
that the shortest wavelength of turbulence must be larger than that gyro-radius of particle. 
A contour plot of $\Phi ( r, \theta )$ by
Eq.(4) is given in Fig.1.

\begin{figure}[tb]
\centering
\includegraphics[width=6.5cm,clip]{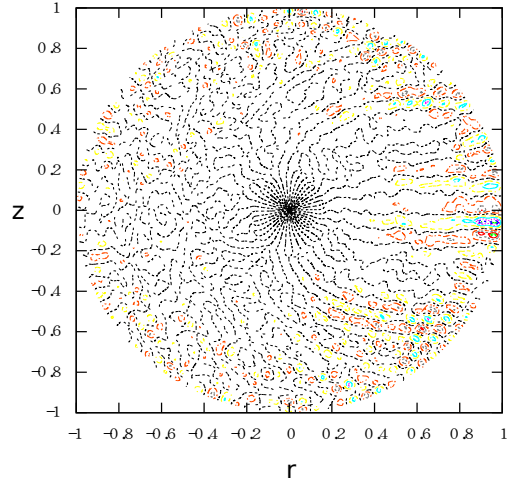}
\caption[Nonlinear ballooning structure on a poloidal plane employed ]
{A contour plot of the electrostatic potential used
in the orbit following calculation given by Eq.(4),
demonstrating nonlinear ballooning structure on a poloidal plane.
Here, $x=(r/a) \cos{\theta}$ and $z=(r/a) \sin{\theta}$. Minor radius at
the wall is given by $a$.}
\end{figure}

\section{Orbit following calculation results}
We now discuss the behavior of passing and 
trapped particles in the presence of background fluctuation. 
In the calculation we use an ITER like profile, $R_0 = 6.2 (m)$, minor radius $a = 2.0 (m)$, and
$B = 5.3 (T)$ at the magnetic axis.
Figure.2 is a Poincar\'{e} plot of guiding center of passing particles 
at a toroidal angle $\zeta = 0$ for 1.5keV passing particles. 
The pitch angle is taken as $\lambda = v_\| /v = 1 $ ($v$ is the total velocity).
In Fig.2, we turned the curvature term off (mirror force and the gradient-B drift terms are
automatically zero when $\lambda=1$ and thus $\mu=0$) for the demonstration purpose, 
but the particles still respond to the background fluctuation through the $E \times B$ term. 
In Fig.2, perturbation modes of $11 \le m \le 29$ with a single $n=10$ mode are incorporated [see Eq.(4)]. 
In the presence of multiple $n$ fluctuation,
the passing particles will be stochastic due to dense formation of guiding center
islands and the overlapping of the islands (see Fig.2 at $r/a < 0.7$)\cite{hua13}.
For the passing particles, the transport is 
much larger for thermal ions than the alpha particles,
because the island width is proportional to $1/\sqrt{v_\|}$\cite{hua13, hua14}.
This is because the net $E \times B$ drift for resonating particles is larger for the slower particles
which spend longer time at the same phase of the driftwave. Note that the latter effect is
different from the idea of {\it orbit averaging}\cite{zha08} which is a crude averaging idea only 
applied for toroidal ripple toroidal modes ``$N$'' (no resonance condition considered)\cite{whi89}.
The finite $\omega_{\star n}$ terms in Eq.(4) gives rise to the radial shift of the resonant
location [positive (negative) $\omega_{\star n}$ for upshift (downshift) in $r$] which match with the
analytical estimation\cite{hua14}. In this paper, we set $\omega_{\star n} = 0$, however.

\begin{figure}[tb]
 \centering
 \includegraphics[width=6.5cm,clip]{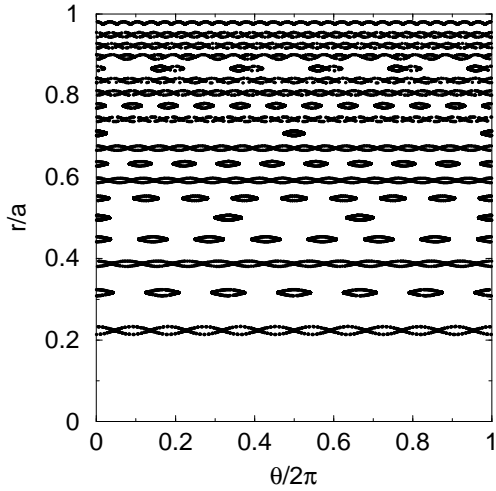}
\caption{Poincar\'{e} plot of guiding centers at a toroidal angle $\zeta = 0$ for $1.5 keV$ deuterium
ions. The pitch angle is taken as $\lambda = v_\| /v = 1 $
and the initial position of the particles are at $(r, \theta, \zeta) = (r_{mn}, 0,0)$.
To demonstrate the guiding center formation the mirror force and the magnetic curvature terms
are off but the $E \times B$ drift effects is kept. The number of island chains
corresponds to the poloidal mode number $m$ while having the same toroidal mode number $n$.
The safety factor profile is given by $q(r) =  1 + 2 (r/a)^2$. Note that
the widths of the guiding center islands are
inverse proportional to the square root of parallel velocities.} 
\end{figure}

In Fig.3, Poincar\'{e} plots of trapped $1.5 keV$ deuterium ions are shown (at $r/a > 0.7$) by keeping all the 
drift terms and the mirror force. Toroidal modes of $10 \le n \le 50$ are incorporated.
Near the banana tips, $v_{\parallel}$ is close to zero so that particle is sensitive to the $E \times B$ drift. 
The driftwave effects, as in the passing particle case, are larger for the lower energy particles.
Note that the $E \times B$ drift itself is not particle's energy dependent but the energy dependence
comes in as a combination of the parallel and perpendicular motion. 
If the parallel velocity is small, they will stay for a long time at the same phase of the perturbation, 
and drift largely in the radial direction. 
The trajectories of trapped particles are not resonant and do not form any islands. 
The trajectories are distorted (invariant tori) without a topological change. 

In Fig.4, Poincar\'{e} plots of trapped $3.5 MeV$ alpha particles are shown. 
Figure 4 demonstrates that the second adiabatic invariant (``$J$'') for 
trapped particles is not easy to break by the background fluctuation
(except for those due to the resonance in the velocity space\cite{hua14}).  

In Fig.3, lower energy trapped particles are seemingly diffusive but 
by if one follows the particle orbits for a longer time,
they can end up in a long term periodic motion. 
The Kolmogorov entropy\cite{nis97,hua14}
for the trapped particles in Fig.3 are estimated which does not necessarily show exponential divergence 
(even for the $1.5 keV$ ions). The radial excursion of trapped particles
is limited in finite radial domains.
In the next section, we investigate collisional effects which can free the trapped
particles from the limited radial domain. 

\begin{figure}[tb]
\centering
\includegraphics[width=6.5cm,clip]{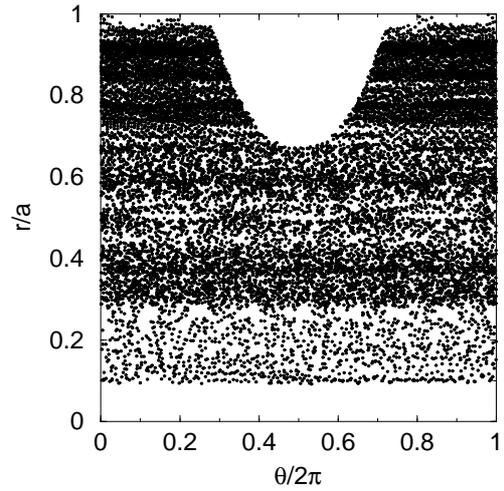}
\caption{Poincar\'{e} plots of guiding centers at a toroidal angle $\zeta = 0$ for $1.5 keV$ deuterium
ions. The pitch angle is taken as $\lambda = v_\| /v = 0.6 $. All the drift terms are kept.
The plot absent region in the center ($r/a \ge 0.7$) corresponds to the strong magnetic
field side of the tokamak. Particles are set initially at $\theta =0$ and equi-distantly
along the $r$ axis.} 
\end{figure}

\begin{figure}[tb]
 \centering
 \includegraphics[width=6.5cm,clip]{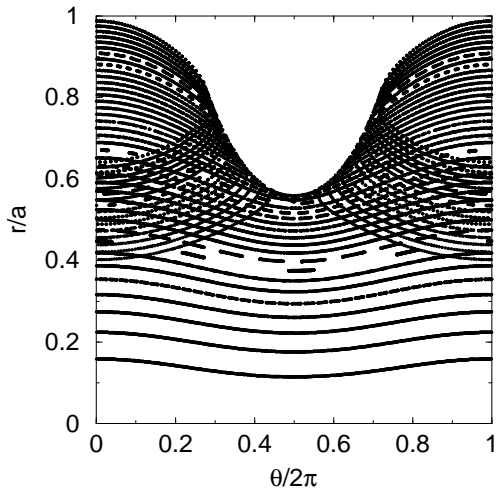}
\caption{Poincar\'{e} plots of guiding centers at a toroidal angle $\zeta = 0$ for $3.5 MeV$ alpha
particles. The rest of the settings is exactly the same with Fig.3.} 
\end{figure}

\section{Coulomb collisional effects on trapped particles}
We employ a Monte Carlo collision operator in the orbit following code. 
The density and the temperature profile for the bulk ions and electrons are
assumed to be $n(r) = n_0 (1-r^2/a^2)$ and $T(r) = T_0 (1-r^2/a^2)$
where $n_0 = 10^{20} (m^{-3})$ and $T_0 = 20 (keV)$. 
The collision effects are given by the energy scattering\cite{nis11} and
the pitch angle scattering\cite{nis08} which obeys the Lorentz collision operator\cite{boz81}.
The collision frequency for the energy scattering by the electrons
is given by the inverse of the energy relaxation time\cite{miy87} while
the collision frequency for the pitch angle scattering (test ion particles against bulk ions) 
is given by the inverse of the perpendicular momentum relaxation time\cite{miy87,nis10}.
The latter collision frequency is changed as a function of the test particle energy.

The diffusion coefficients $D$ in MKS units
are obtained from estimating the second order cumulant\cite{nis97}.
Figure 5 shows the the diffusion coefficients versus test ion energy, for (black) trapped particles
without collisions but with the turbulent background, 
and (red) trapped particles with collisions and the turbulence. 
In estimating the diffusion,
$D = (1/N_p) \sum_{i=1}^{N_p} (r - r_0)^2/2 t_s $ ($r_0$ is the initial radial location
and $t_s$ is the sampling time), we have set $N_p = 448$ test particles separated 
equi-distantly in $0 \le \zeta \le 2 \pi$  at $r/a=0.5$ (where the background bulk
temperature is $15keV$) and $ \theta = 0$
with an initial pitch angle $\lambda = 0.5$, which are all trapped particles.
To subtract the finite banana width effects (which is large for high energy particles) 
from the diffusion estimation, 
radial displacement is measured when the banana motions cross $\theta = 0$
after each bounce motion.

In the guiding center calculation, all the drift terms are incorporated.
The toroidal modes of $10 \le n \le 50$ are employed.
The maximum amplitude of the electrostatic fluctuation is given by
$|e \Phi/T_0| = 10^{-3}$ where $e$ is the unit charge.

As shown in Fig.5, the diffusion level decreases as the energy
increase for both the curves (the same trend is found for the passing particles\cite{hua14}).
The collisional effects are visible toward the lower energy side, $E_{test} \le 50 keV$.
Note that both the abscissa and the ordinate are in the logarithmic scale.
Although the collision frequencies are very small in the high temperature collisionless
plasmas, it is suggested that the collisions (pitch angle scattering per se) 
can possibly remove the trapped particles from a long term periodic motion.

\begin{figure}[tb]
\centering
\includegraphics[width=7.5cm,clip]{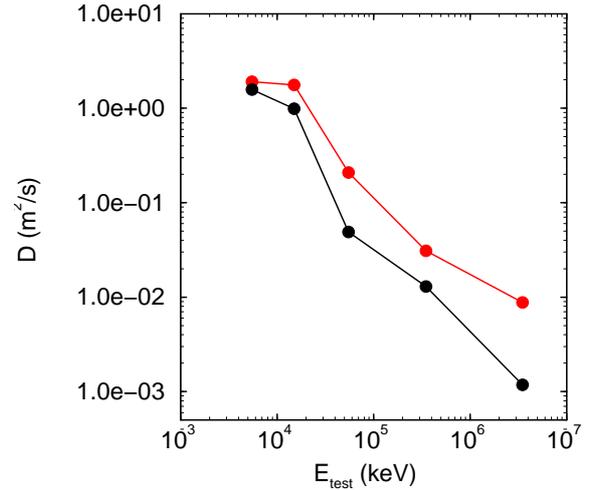}
\caption{Diffusion coefficients $D$ versus test ion energy $E_{test}$; (black curve) trapped particles
without collisions but with the turbulent background, 
and (red curve) trapped particles with collisions and the 
turbulence. Initial positions of the test particles are given by $r/a=0.5$, $ \theta = 0$,
and $0 \le \zeta \le 2 \pi$.
Pitch angle of $\lambda = 0.5$ taken for all the cases.} 
\end{figure}

\section{Summary and discussions}
In this work, we have studied high energy particle transport in the presence of
toroidal driftwave turbulence. 
To model more realistic trapped particle transport, 
energy and pitch angle scattering are incorporated.
It is suggested that small collisional effects can change the transport process of the trapped particles.

The authors would like to thank discussions with Dr. S. Satake of National Institute for Fusion Science. 
One of the authors YN would like to thank discussions with Dr. M. Azumi.
This work is supported by National Cheng Kung University
Top University Project and by National Science Council of Taiwan,
NSC 100-2112-M-006-021-MY3.

\end{normalsize}

\end{document}